\documentclass[aps,prl,twocolumn,10pt,floatfix]{revtex4-1}
\usepackage{amssymb,amsmath,amsfonts,amsbsy,epsfig}

\newcommand{\be}{\begin{equation}}
\newcommand{\ee}{\end{equation}}
\newcommand{\bea}{\begin{eqnarray}}
\newcommand{\eea}{\end{eqnarray}}
\newcommand{\nn}{\nonumber}

\begin{document}

\title{Sound Speed of Primordial Fluctuations in Supergravity Inflation }
\date{October 11, 2016}

\author{
Alexander Hetz and Gonzalo A. Palma
}

\affiliation{
Grupo de Cosmolog\'ia y Astrof\'isica Te\'orica, Departamento de F\'{i}sica, FCFM, \mbox{Universidad de Chile}, Blanco Encalada 2008, Santiago, Chile.}

\begin{abstract} 

We study the realization of slow-roll inflation in $\mathcal N = 1$ supergravities where inflation is the result of the evolution of a single chiral field. When there is only one flat direction in field space, it is possible to derive a single-field effective field theory parametrized by the sound speed $c_s$ at which curvature perturbations propagate during inflation. The value of $c_s$ is determined by the rate of bend of the inflationary path resulting from the shape of the $F$-term potential. We show that $c_s$ must respect an inequality that involves the curvature tensor of the K\"ahler manifold underlying supergravity, and the ratio $M/H$ between the mass $M$ of fluctuations ortogonal to the inflationary path, and the Hubble expansion rate $H$. This inequality provides a powerful link between observational constraints on primordial non-Gaussianity and information about the $\mathcal N = 1$ supergravity responsible for inflation. In particular, the inequality does not allow for suppressed values of $c_s$ (values smaller than $c_s \sim 0.4$) unless (a) the ratio $M/H$ is of order $1$ or smaller, and (b) the fluctuations of mass $M$ affect the propagation of curvature perturbations by inducing on them a nonlinear dispersion relation during horizon crossing. Therefore, if large non-Gaussianity is observed, supergravity models of inflation would be severely constrained.

\end{abstract}

\maketitle

It is unlikely that observations will ever allow us to pin down the fundamental theory underlying inflation~\cite{Guth:1980zm, Linde:1981mu, Albrecht:1982wi, Starobinsky:1980te, Mukhanov:1981xt}. Instead, we might have to conform ourselves with gaining general insights into the origin of cosmological perturbations with the help of the effective field theory (EFT) approach to inflation~\cite{Cheung:2007st} (see also Ref.~\cite{Weinberg:2008hq}). This consists of a powerful model-independent framework to study the evolution of curvature perturbations $\mathcal R$ during inflation, valid at energies around the Hubble expansion rate $H$. With the help of symmetry arguments, it is possible to derive the most general class of Lagrangians governing the dynamics of fluctuations in a Friedman-Robertson-Walker (FRW) background, in terms of a Goldstone boson field $\pi = - {\mathcal R} / H$, given by an operator expansion with nonlinearly related coefficients parametrizing physics beyond the canonical single-field paradigm. These coefficients, among which the sound speed $c_s$ plays a prominent role, are determined by the ultraviolet (UV) complete theory underlying inflation. To quadratic order, the Lagrangian is given by~\cite{Cheung:2007st}
\be
{\mathcal L}^{(2)}_\pi =  M_{\rm Pl}^2 | \dot H | \left[ c_s^{-2} \dot \pi^2 - a^{-2} ( \nabla \pi )^2  \right] , \label{L-pi} 
\ee
where $M_{\rm Pl}$ is the reduced Planck mass and $a$ is the scale factor. $c_s$ also defines the strength of cubic operators that may lead to primordial non-Gaussianity~\cite{Chen:2006nt}. For instance, if $c_s < 1$ (with $|\dot c_s| \ll H c_s$), then equilateral and orthogonal non-Gaussianity are produced, with $f_{\rm NL}$ parameters given by~\cite{Senatore:2009gt}
\bea
f_{\rm NL}^{\rm equil} = - (c_s^{-2} - 1) ( 0.275 + 0.078 c_s^2 + 0.52 \tilde c_3), \label{fnl-1} \\
f_{\rm NL}^{\rm ortho} =  (c_s^{-2} - 1) ( 0.0159 - 0.0167 c_s^2 + 0.11 \tilde c_3), \label{fnl-2}
\eea
where $\tilde c_3$ is a parameter that arises at third order~\cite{Senatore:2009gt} and it is determined by the UV-complete theory~\cite{Note1}. It has been conjectured~\cite{Baumann:2015nta} that every parameter of the EFT expansion, such as $\tilde c_3$, is proportional to $(1 - c_s^2)$. If so, single field slow-roll inflation would be the only single field theory such that $c_s = 1$.  

Current constraints on primordial non-Gaussianity give $c_s \geq 0.024$ (95\%~C.L.) after marginalizing over $\tilde c_3$~\cite{Ade:2015ava}. Future measurements will further constrain the coefficients of the EFT Lagrangian, narrowing down the class of fundamental theories that can host inflation together with the standard model of particle physics. Thus, it is important to understand how fundamental theories map into the EFT expansion, especially in the case of theories such as supergravity and string theory, characterized for having a large number of scalar fields representing moduli that must remain massive during inflation.

Given that supergravities have several scalar fields, it is conceivable that the inflationary path experienced bends in field space. In fact, it is well understood that, regardless of how massive fields normal to the path are, these bends affect the dynamics of $\pi$ by inducing $c_s < 1$~\cite{Tolley:2009fg, Achucarro:2010da}. For instance, in two-field models one has~\cite{Achucarro:2010da, Achucarro:2012sm, Achucarro:2012yr}
\be
c_s = \left( 1 + 4 \Omega^2 / M^2 \right)^{-1/2} , \label{cs-Omega}
\ee
where $\Omega$ is the local angular velocity describing the bend, and $M$ is the mass of fields orthogonal to the inflationary path. In this Letter we will deduce an inequality that any single field EFT derived from $\mathcal N=1$ supergravity must satisfy. It involves $c_s$, $H$, and $M$, and it is given by
\begin{equation} 
\left( c_s^{-2}  - 1 \right)   M^2 / H^2  \leq \mathcal O ( 1 ) , \label{main-result}
\end{equation}
where $\mathcal O ( 1 )$ represents an order $1$ number determined by the geometry of the K\"ahler manifold underlying the supergravity of interest. As we shall see, to ensure that the horizon exit is properly accounted for by Eq.~(\ref{L-pi}), one requires that $(1 - c_s^2)^2 H^2/ M^2 c_s^2  \ll 1$. If not, the dispersion relation describing the propagation of $\pi$ during horizon crossing becomes nonlinear, due to new operators in Eq.~(\ref{L-pi}) with spatial gradients induced by the massive field normal to the path. This, together with Eq.~(\ref{main-result}) implies that $c_s$ cannot be much smaller than $1$ (and supergravity models cannot produce large primordial non-Gaussianity) unless the horizon crossing is described by an EFT with a nonlinear dispersion relation~\cite{Baumann:2011su, Gwyn:2012mw}. In this case, the dynamics of fluctuations would depart considerably from the standard single-field EFT description of Eq.~(\ref{L-pi}) used to connect models with observables. 

\begin{figure}[t!]
\includegraphics[scale=0.55]{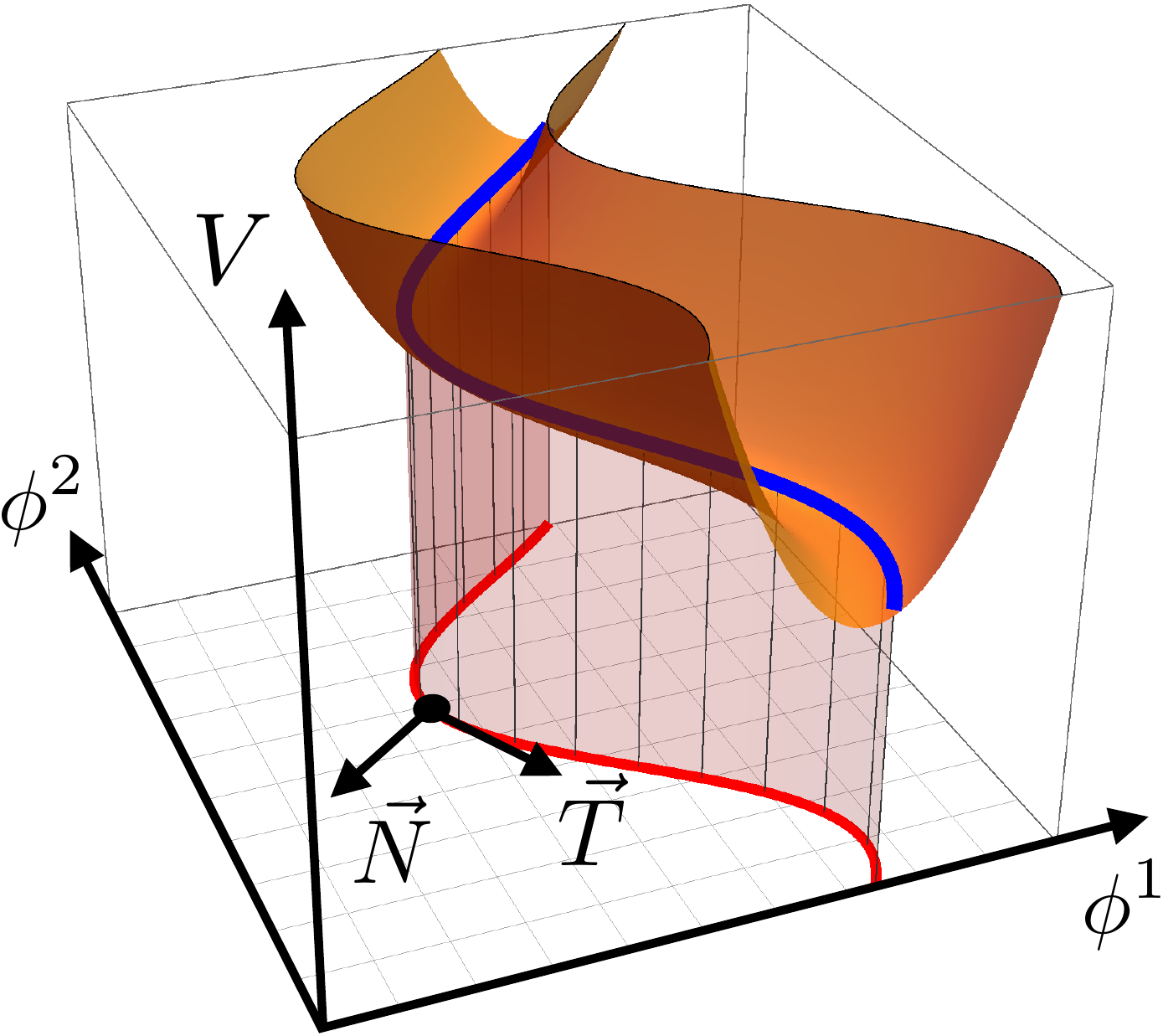}
\caption{An example of a path forced to have bends as it meanders down a two-field potential $V$.}
\label{fig:fig_01}
\end{figure}

Before deducing the inequality, let us review some basics about multifield inflation. We consider a general action $S_{\rm tot} = \int \! \sqrt{-g} \, {\mathcal L}$ describing an arbitrary number of real scalar fields $\phi^a (t, {\bf x})$, $a=1, \ldots , N$, minimally coupled to gravity, with a Lagrangian
\be
{\mathcal L} =  \frac{M_{\rm Pl}^2}{2} R + \frac{1}{2}\gamma_{ab} \, g^{\mu\nu}\partial_{\mu}\phi^a\partial_{\nu}\phi^b - V(\phi) , \label{lagrangian}
\ee
where $R$ is the Ricci scalar constructed from the metric $g_{\mu \nu}$, $\gamma_{ab}$ is a $\sigma$-model metric that depends on the fields, and $V (\phi)$ is the scalar potential. From now on, we work in units where $M_{\rm Pl} = 1$. In the case of $\mathcal N = 1$ supergravity, $\gamma_{ab}$ and $V$ are uniquely determined by the function $G = K +\ln |W|^2$, where $K$ is the K\"ahler potential, and $W$ the superpotential. We will come back to supergravity in a moment. To describe the background, we consider an FRW spacetime with a metric $ds^2 = - dt^2 + a^2(t) d {\bf x}^2$, where $a(t)$ is the scale factor and ${\bf x}$ are comoving coordinates. The expansion rate is $H= \dot a /a$ and it is given by the Friedman equation $3 H^2  = \dot \phi_0^2 / 2+ V(\phi)$, where $\dot \phi_0^2 \equiv \gamma_{a b} \dot \phi^a_0 \dot \phi^b_0$. The equations of motion for the homogeneous background fields $\phi^a_0(t)$ derived from Eq.~(\ref{lagrangian}) are
\be
D_t \dot \phi_0^a + 3 H \dot \phi_0^a + \gamma^{a b} V_b = 0,  \label{eq_mot}
\ee
where $V_b = \partial_b V = \partial V / \partial \phi^b$. Furthermore, $D_t$ is a covariant time derivative whose action on a given vector $A^a$ is such that $D_t A^a \equiv \dot A^a + \Gamma^{a}_{bc} A^b \dot \phi_0^c$, where $\Gamma^{a}_{bc}$ are the Christoffel symbols derived from $\gamma_{ab}$. The inflationary perturbations are defined as $\delta \phi^a (t , {\bf x}) = \phi^a (t, {\bf x}) - \phi^a_0 (t)$. To study them, it is useful to define vectors tangent and normal to the path $\phi^a_0(t)$, respectively, given by
\be
T^a \equiv  \dot \phi_0^a  /  \dot \phi_0 , \qquad N^a  \equiv -  D_t T^a /  |D_t T| . \label{def-unit-vectors}
\ee
Figure~\ref{fig:fig_01} illustrates an example where inflation is driven by a two-field potential. 
Fluctuations along the direction $T^a$ give us curvature perturbations as ${\mathcal R} = - H T_a \delta \phi^a / \dot \phi_0$, whereas those along $N^a$ correspond to isocurvature perturbations~\cite{Gordon:2000hv, GrootNibbelink:2001qt}. $T^a$ and $N^a$ allow us to define the angular velocity $\Omega$ via $D_t T^a  \equiv - \Omega N^a$ [then Eq.~(\ref{def-unit-vectors}) tells us that $\Omega \geqslant 0$]. From Eq.~(\ref{eq_mot}) it follows that $V_a = V_\phi T_a + V_N N_a$, where $V_\phi = T^a V_a$ and $V_N = N^a V_a$. As a result, projecting Eq.~(\ref{eq_mot}) along $T^a$ one finds $\ddot \phi_0 + 3 H \dot \phi_0 + V_\phi = 0$, which resembles the equation of motion of a single scalar field. On the other hand, projecting Eq.~(\ref{eq_mot}) along $N^a$ one obtains,
\be
\Omega =  V_N / \dot \phi_0 . \label{Omega-V_N}
\ee
This equation reveals that whenever the path is subjected to a bend, it is pushed towards the outer wall of the potential. $\Omega$ plays a crucial role in the dynamics of fluctuations, as it couples together curvature and isocurvature modes~\cite{GrootNibbelink:2001qt}. To be consistent with the view that there must be a shift symmetry ensuring a flat direction in the potential, in this work, we assume that $\delta_{\Omega} \equiv \dot \Omega / H \Omega$ is suppressed (steady bends), implying that $c_s$ varies slowly: $|\dot c_s| \ll H c_s$. One may consider sharp bends, in which case features are produced in the spectra~\cite{Achucarro:2010da}, and Eqs.~(\ref{fnl-1}) and~(\ref{fnl-2}) are not sufficient to constrain $c_s$.

In order to characterize the evolution of $H$ during inflation it is useful to define the slow-roll parameters:
\be
\epsilon \equiv -  \dot H / H^2 = \dot \phi_0^2/2 H^2, \qquad \eta \equiv  \dot \epsilon / H \epsilon . \label{def_slow-roll}
\ee
Slow-roll inflation corresponds to the regime where the expansion rate $H$ evolves very slowly for a long enough time, requiring that $\epsilon \ll 1$ and $|\eta | \ll 1$. In this regime, one finds $3 H^2 \simeq V$, and $\epsilon \simeq  V_{\phi}^2 / 2 V^2$. On the other hand, it is also customary to introduce a matrix proportional to the Hessian of the potential $V$:
\be
N_{a b} \equiv   \nabla_a V_b / V  .
\ee
 This matrix contains information about the shape of the potential, and it is sometimes used to define slow-roll parameters alternative to those of Eq.~(\ref{def_slow-roll}). These are~\cite{Burgess:2004kv}
\be
\epsilon_V \equiv  \frac{1}{2} \frac{V_a V^a}{V^2}  , \qquad \eta_{V} \equiv \textrm{min. eigenvalue} \, \{  N \} .
\ee
However, there is a problem with these definitions: Both $\epsilon_V$ and $\eta_V$ may attain large values even with $\epsilon, |\eta | \ll 1$. First, $\epsilon_V$  is not necessarily small because $T^a$ and the direction of steepest descent of the potential ($\propto V_a$) do not coincide in a bend. To see this, we may combine $\epsilon \simeq  V_{\phi}^2 / 2 V^2$ with Eqs.~(\ref{Omega-V_N}) and (\ref{def_slow-roll}) to find that $\epsilon_V$ is given by
\be
\epsilon_V = \epsilon \left( 1 +  \Omega^2 / 9 H^2 \right) .
\ee
This shows that $\epsilon_V$ may attain values of order $1$ without necessarily violating slow roll. Second, $\eta_V$ is not necessarily small because $T^a$ and the eigenvector associated to $\eta_V$ (the smallest eigenvalue of $N_{ab}$) do not coincide in a bend (that is, $\eta_{V} \neq N_{a b} T^a T^b$ when $\Omega \neq 0$). To show this, let us compute $\eta_V$ in the case of two-field models. In this case, we may define a unit vector $h^a \equiv \cos \alpha \,T^a + \sin \alpha \, N^a$ and minimize the contraction $h^a h^b \nabla_a V_b / V$ by tuning $\alpha$, to obtain back $\eta_V$. To perform the contraction we need to be aware of the following identities~\cite{Achucarro:2010da}:
\bea
T^a T^b \nabla_a V_{b} &=& 3 H^2 (2 \epsilon - \eta /2)  + \Omega^2 ,  \label{TT}  \\ 
N^a T^b \nabla_a V_{b} &=& \Omega H  ( 3 + \delta_{\Omega}+ \eta - 2 \epsilon )  ,  \label{NT} \\
N^a N^b \nabla_a V_{b} &=& M^2 + \Omega^2 - \epsilon H^2 \mathbb{R} ,  \label{NN}
\eea
where $\delta_{\Omega} = \dot \Omega / H \Omega$, and $\mathbb{R} = R_{a b c d} T^a N^b T^c N^d$ is the Riemann tensor of the scalar field space projected along $T^a$ and $N^b$. $M$ is the effective mass of the perturbation along the direction $N^a$, and it is the same quantity appearing in Eq.~(\ref{cs-Omega}). Given that we are assuming slow roll, we disregard $\delta_{\Omega}+ \eta - 2 \epsilon $ in Eq.~(\ref{NT}) before minimizing $h^a h^b \nabla_a V_b / V$. In addition, we may assume that $\epsilon H^2 \mathbb{R}$  is small compared to $M^2 + \Omega^2$, which is generally true in supergravity models of inflation~\cite{Ellis:2013xoa, Ellis:2013nxa, Carrasco:2015rva, Carrasco:2015pla}. Then, we find
\bea
\eta_V = \frac{1}{12} \Bigg(   (c_s^{-2} - 1 ) \frac{M^2}{H^2}  + 2 \frac{M^2}{H^2}  +  3  \left[ 4 \epsilon - \eta \right] \qquad \quad \nn \\ 
 -  2  \sqrt{  \left(   \frac{M^2}{H^2} -  \frac{3}{2}  \left[ 4 \epsilon - \eta \right] \right)^2   +  9 (c_s^{-2} - 1 ) \frac{M^2}{H^2}    } \Bigg),  \quad \label{eta_V_two_fields}
\eea
where we used Eq.~(\ref{cs-Omega}) to introduce $c_s$. Now, the crucial point to appreciate here is that it is possible to have $\Omega \gg M $ (and therefore $c_s^{2} \ll 1$) without violating the conditions $\epsilon , |\eta | \ll 1$. If this is the case, the misalignment is large, and $\eta_V$ acquires sizable values. On the contrary, if $\Omega = 0$, one recovers a suppressed value given by $\eta_V = (4 \epsilon - \eta) / 2$. Hence, $\epsilon_V$ and $\eta_V$ are not legitimate slow-roll parameters unless the multifield path remains straight ($\Omega = 0$).

Let us now deduce the inequality~(\ref{main-result}), valid for the scalar sector of $\mathcal N = 1$ supergravity. Before considering the case of a single chiral field, let us consider the general situation. Here, the target space corresponds to a K\"ahler manifold spanned by complex scalar fields $\Phi^i$. The $F$-tern potential is given by $V = e^G (K^{i \bar \jmath} G_i G_{\bar \jmath} - 3)$, where $G_i = \partial_i G$ and $K_{i \bar \jmath} = \partial_i \partial_{\bar \jmath} G$ is the K\"ahler metric (derivatives are taken with respect to $\Phi^i$ and their complex conjugate $\Phi^{\bar \imath}$). One may now define a Hessian matrix $\nabla_i V_{\bar \jmath}$. Then, because $\eta_V$ is the minimum eigenvalue of $N_{ab}$, one necessarily has~\cite{Covi:2008cn}
\be
\eta_{V} \leq   f^i f^{\bar \jmath} \nabla_i V_{\bar \jmath} / V , \label{ineq-f}
\ee
where $f^i$ is an arbitrary unit complex vector. It turns out that if $f^i = G^i / \sqrt{ G^j G_j }$, then the right-hand side (rhs) of Eq.~(\ref{ineq-f}) becomes independent of second derivatives of the superpotential $W$~\cite{GomezReino:2006dk, Covi:2008ea}, and reduces to~\cite{Covi:2008cn}
\bea
 \frac{\nabla_i V_{\bar \jmath} }{V } f^i f^{\bar \jmath} \leq -  \frac{2}{3} + \frac{1+\gamma}{\gamma} \left[ \frac{2}{3}  -  \mathbb{R}(f) \right]  \qquad \nn \\
+ \frac{\epsilon_V}{1+\gamma}  +\frac{4}{\sqrt{3}}\frac{ \sqrt{ \epsilon_V }}{\sqrt{1+\gamma}} , \qquad \label{ineq_Vff}
\eea
where $\gamma \equiv H^2 / m_{3/2}^2$ ($m_{3/2} = e^{G/2}$ is the gravitino mass), and $\mathbb{R}(f)  \equiv R_{i \bar \jmath k \bar l} f^i f^{\bar \jmath} f^k f^{\bar l}$ is the Riemann tensor computed out of $K_{i \bar \jmath}$ projected along the complex vector $f^i$. 

We now have all the necessary elements to deduce the desired inequality: If the supergravity has only one chiral field $\Phi$, then the scalar field target space is spanned by two real scalar fields, implying that $\eta_V$ in the left-hand side (lhs) of Eq.~(\ref{ineq-f}) is precisely given by Eq.~(\ref{eta_V_two_fields}). Then, putting together Eqs.~(\ref{eta_V_two_fields}) and~(\ref{ineq-f}) we deduce the desired inequality. However, in order to explicitly write down a simple version of it, it is useful to anticipate three relevant regimes: First, if there are no bends, $\eta_V = (4 \epsilon - \eta) / 2$, and one recovers the bound deduced in Ref.~\cite{Covi:2008cn} (see also Ref.~\cite{Borghese:2012yu}), informing us whether a given supergravity can produce inflation without bends. Second, if $(c_s^{-2} - 1) M^2 / H^2$ is of order $\mathcal O(\epsilon , \eta)$, then $\Omega^2 \ll H^2$, and $c_s$ is close to $1$. Then, $\eta_V$ is of order $\mathcal O(\epsilon , \eta)$ and the inequality emerging from Eq.~(\ref{ineq-f}) does not give us interesting information beyond that of Ref.~\cite{Covi:2008cn}. Third, if $(c_s^{-2} - 1) M^2 / H^2 \gg  | \mathcal O(\epsilon , \eta) |$ we may neglect the slow-roll parameters appearing in Eqs.~(\ref{eta_V_two_fields}) and~(\ref{ineq_Vff}) to obtain our main result:
\bea
 \frac{1}{6} \frac{M^2}{H^2} \left(   \frac{c_s^{-2} - 1}{2}   + 1    -   \sqrt{  1    +  9 (c_s^{-2} - 1 ) \frac{H^2}{M^2}    } \right) \nn\\
 \leq  \frac{1+\gamma}{\gamma} \left[ \frac{2}{3}  -  \mathbb{R}(f) \right] -  \frac{2}{3} . \label{main-result-full}
\eea
This inequality puts together the sound speed $c_s$, the ratio $M/H$, and additional information on the curvature of the K\"ahler manifold.

Let us now analyze the content of Eq.~(\ref{main-result-full}), which assumes that $\epsilon, |\eta| \ll 1$~\cite{Note2}. To start with, notice that the smallest attainable value of the lhs of (\ref{main-result-full}) is $- 3/4$. This means that when the inflationary path is subjected to strong bends, it is possible to have $M^2 > 0$ simultaneously with a large and negative value of $\eta_V$. Next, the rhs of eq.~(\ref{main-result-full}) may acquire any desired value, depending on the specific K\"ahler potential under consideration. Nevertheless, we expect this value to be of order $1$. First, the natural value of the curvature term is $\mathbb{R}(f) \simeq \mathcal O (1)$. For instance, canonical models are characterized by $\mathbb{R}(f)  = 0$ whereas single-chiral-field no-scale models satisfy $\mathbb{R}(f) = 2/3$~\cite{Cremmer:1983bf}. Second, there are two classes of models of supergravity inflation depending on the value of $\gamma = H^2 / m_{3/2}^2$. On the one hand, we expect the value of $H$ during inflation to be of order $10^{14}$GeV or smaller, but not too much smaller (in order to have a tensor to scalar ratio $r \lesssim 0.1$). On the other hand, we expect the value of $m_{3/2}$ at the end of inflation to be of order $1$TeV or larger, but not too much larger (in order to have a soft SUSY breaking scale close to the electroweak symmetry breaking scale). Hence, in models where $m_{3/2}$ did not evolve significantly during inflation, one expects $\gamma \gg 1$. To be conservative, in the present Letter we assume $\gamma \gtrsim 1$, implying that the rhs of Eq.~(\ref{main-result-full}) is of order $1$. There are string inspired models, such as large volume scenarios~\cite{Conlon:2008cj}, where $\gamma \ll 1$. However, in these models one finds $2/3  -  \mathbb{R}(f) \sim \gamma$, implying that the rhs eq.~(\ref{main-result-full}) continues to be of order 1~\cite{Note3}.

\begin{figure}[t!]
\includegraphics[scale=0.43]{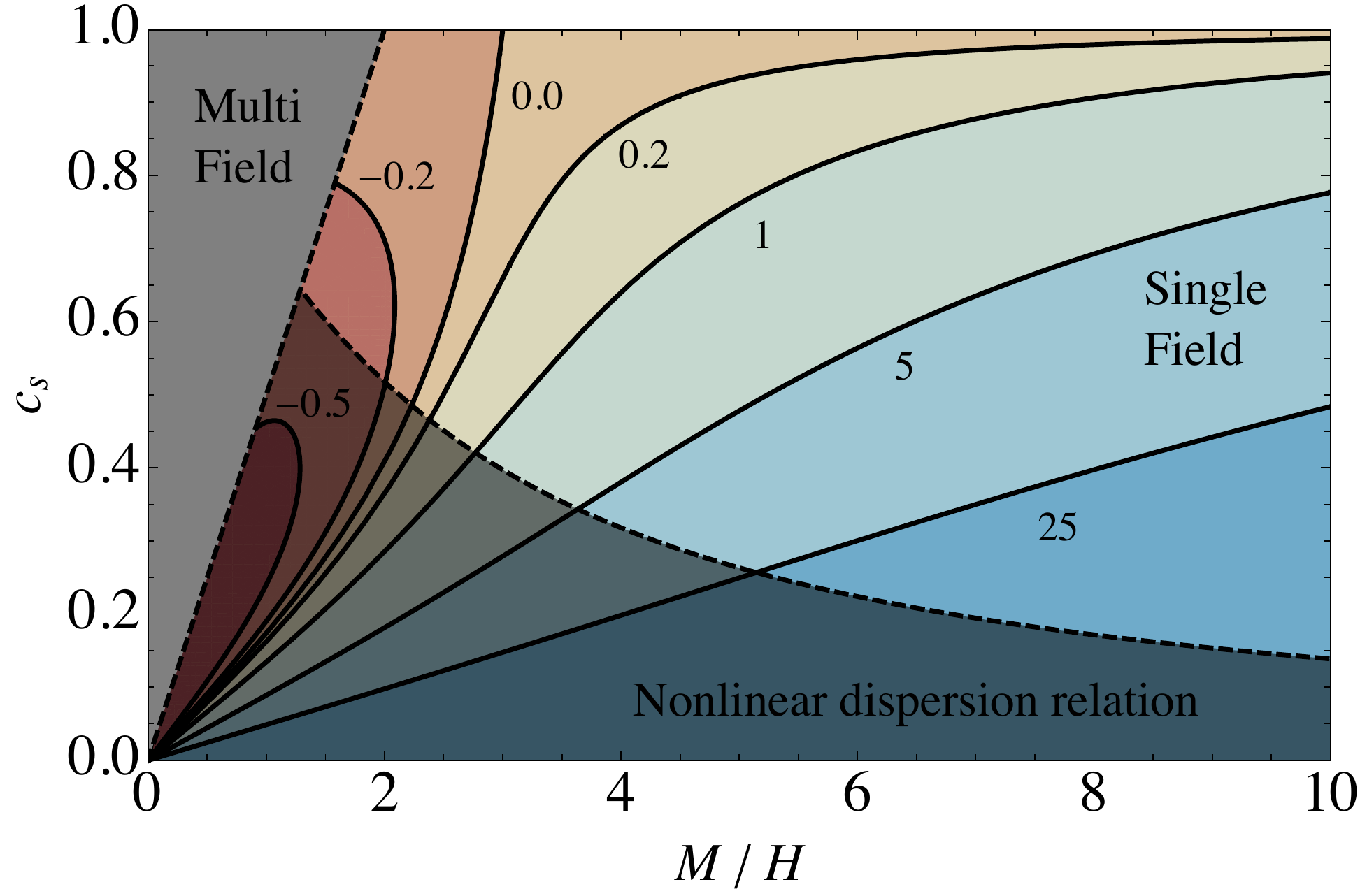}
\caption{The parameter space $M/H$ vs $c_s$. The inequality is satisfied at the regions bounded by the solid lines, opposite to the labels indicating benchmark values for the rhs of Eq.~(\ref{main-result-full}).}
\label{fig:fig_02}
\end{figure}

It is important to consider in our analysis the range of parameters for which the two-field system admits an EFT of the form~(\ref{L-pi}). This will depend on the values of the ratios $M/H$ and $\Omega / M$ (or equivalently $c_s$). We plot these ranges in Fig.~\ref{fig:fig_02}: First, the dynamics of fluctuations is described by a single degree of freedom as long as the Goldstone boson's energy $\omega$ satisfies $\omega^2 \ll M^2 / c_s^2$~\cite{Achucarro:2012yr}. Then, requiring that horizon crossing $\omega \simeq H$ happens within this regime, we obtain the bound $H^2 \ll M^2 / c_s^2$. This bound defines the region at the right of the ``Multi Field" area in Fig.~\ref{fig:fig_02}, which, for reference, is limited by the curve $H^2 = 0.5 M^2 / c_s^2$. Within this region there are two additional subregimes: If $\omega^2 \ll M^2 c_s^2 / (1 - c_s^2)^2$ the dispersion relation is of the form $\omega (k) = c_s k$ and the theory is described by Eq.~(\ref{L-pi}). For energies $\omega^2 \simeq M^2 c_s^2 / (1 - c_s^2)^2$ or larger, the spatial gradients in the kinetic term of the massive field (normal to the path) dominate over its mass ($k^2 \gg M^2$), inducing new operators in Eq.~(\ref{L-pi}) that make the dispersion relation $\omega (k)$ for the Goldstone boson $\pi$ nonlinear~\cite{Achucarro:2012yr, Baumann:2011su, Gwyn:2012mw}. In this case, the evolution of $\pi$ is determined by an EFT that departs significantly from~(\ref{L-pi}), implying different relations between observables and the parameters of the theory. Thus, we deduce that the EFT~(\ref{L-pi}) is valid as long as~\cite{Achucarro:2012yr}:
\be
(1 - c_s^2)^2 H^2/ M^2 c_s^2  \ll  1 . \label{EFT-validity}
\ee 
The region satisfying this bound corresponds to the area above the dashed curve satisfying $(1 - c_s^2)^2 H^2/ M^2 c_s^2  =  0.5$, separating the domains ``Single Field" and ``Nonlinear dispersion relation" in Fig.~\ref{fig:fig_02}. In addition, the inequality~(\ref{main-result-full}) is satisfied at the regions bounded by the solid lines, opposite to the labels indicating benchmark values for the rhs of Eq.~(\ref{main-result-full}). 
We may further simplify the form of our inequality: Because of Eq.~(\ref{EFT-validity}) and the fact that the rhs of Eq.~(\ref{main-result-full}) is of order $1$, we may take $9 (c_s^{-2} - 1) \ll M^2 / H^2$ and Taylor expand the square root in Eq.~(\ref{main-result-full}) to recover Eq.~(\ref{main-result}). As a result, for supergravity inflation to be described by Eq.~(\ref{L-pi}),  $c_s$ cannot be much smaller than unity. Indeed, in Fig.~\ref{fig:fig_02} we see that $c_s$ cannot be smaller than $c_s \simeq 0.4$ when the rhs of the inequality is $1$.

In the case of supergravities with many chiral fields, the qualitative content of this result will remain valid. In those cases the rhs of Eq.~(\ref{main-result-full}) will not change, and we only need to worry about the lhs. There, $c_s$ will have a similar dependence on the masses of heavy fields to that of Eq.~(\ref{cs-Omega})~\cite{Cespedes:2013rda}. Moreover, the form of $\eta_V$ will change due to the new matrix entries determined by the masses of the extra fields. However, it turns out that these contributions do not modify $\eta_V$ importantly if the masses of the extra fields are of the same order than that along $N^a$. In particular, in many models of supergravity there is only one chiral field $\Phi$ that evolves during inflation, while the rest remain stabilized, having the role of breaking SUSY~\cite{Baumann:2011nk}. In those types of models our inequality remains exactly valid. This is because here the only evolving fields are the two real scalar fields entering $\Phi$, giving us back the same expression for $\eta_V$ of Eq.~(\ref{eta_V_two_fields}), deduced under the assumption that the inflationary path was embedded in a two-field manifold.

In conclusion, we have derived a direct link between observables and $\mathcal N = 1$ supergravity inflation. By constraining non-Gaussianity, we are able to infer nontrivial information about the supergravity responsible for inflation. If inflation is described by the EFT~(\ref{L-pi}), then $c_s$ cannot be suppressed, and non-Gaussianity must be such that $f_{\rm NL}^{\rm eq, orth} \lesssim \mathcal O (1)$ ($c_s \gtrsim 0.4$). On the contrary, if observations ever confirm sizable non-Gaussianity $f_{\rm NL}^{\rm eq, orth} \gtrsim \mathcal O (1)$ then the supergravity responsible for inflation must have had nontrivial properties pertaining to the evolution of perturbations during horizon crossing: (a) the ratio $M/H$ was of order one or smaller, and (b) the propagation of curvature perturbations during horizon crossing was characterized by a nonlinear dispersion relation as a result of their interaction with the field normal to the trajectory. In such a case, we would need to understand the generation of primordial perturbations within a regime that so far has not been studied in the context of supergravity inflation. Of course, our bound $c_s \gtrsim 0.4$ must be taken with some caution, as the rhs of Eq.~(\ref{main-result-full}) may acquire large or small values depending on the specific model under consideration. Notwithstanding, our analysis gives valuable information about the connection between supergravity and observables, that could be further complemented with constraints on $\tilde c_3$ (which also depends on the K\"ahler geometry).

\begin{acknowledgments}

We are grateful to Ana Ach\'ucarro and Daniel Green for useful comments and discussions that helped to improve the first version of this manuscript. This work was supported by the Fondecyt Project No. 1130777 (G.A.P.), the Conicyt Anillo project ACT1122 (G.A.P), and the Conicyt Fellowship CONICYT-PCHA/Mag\'ister Nacional/2014-66185 (A.H.).\\

\end{acknowledgments}

\end{document}